\documentclass{article}%
\usepackage{amsfonts}
\usepackage{amsmath}%
\usepackage{amssymb}%
\usepackage{graphicx}
\providecommand{\U}[1]{\protect\rule{.1in}{.1in}}
\newtheorem{theorem}{Theorem}
\newtheorem{acknowledgement}[theorem]{Acknowledgement}

\begin{document}

\title{Siewert solutions of transcendental equations, generalized Lambert functions
and physical applications }
\author{Victor Barsan\\National Institute of Physics and Nuclear Engineering (NIPNE),\\Str. Reactorului 30, 077125 Magurele, Romania}
\maketitle

\begin{abstract}
We review the exact solutions of several transcendental equations, obtained by
Siewert and his co-workers, in the '70s. Some of them are expressed in terms
of the generalized Lambert functions, recently studied by Mez\"{o}, Baricz and
Mugnaini. For some others, precise analytical approximations are obtained. In
two cases, the asymptotic form of Siewert's solutions are written as Wright
$\omega$ functions.

\end{abstract}

\section{Introduction}

In a series of papers published between 1972 and 1976 \cite{[S50]} /S50,
\cite{[S52]} /S52, \cite{[S57]} /S57, \cite{[S53]} /S53, \cite{[S56]} /S56,
\cite{[S59]} /S59, \cite{[S62]} /S62, \cite{[S63]} /S63, \cite{[S68]} /S68,
\cite{[S71]} /S71, \cite{[S80]} /S80, \cite{[S89]} /S89, \cite{[S100]} /S100,
\cite{[S108]} /S108, Siewert and his co-workers - Burniston (for \cite{[S50]},
\cite{[S52]}, \cite{[S53]}, \cite{[S59]}, \cite{[S62]}, \cite{[S71]},
\cite{[S80]}), Phelps III (for \cite{[S100]}, \cite{[S108]}), Essig (for
\cite{[S56]}), Dogget (for \cite{[S71]}) and Burkart (for \cite{[S63]}) -
studied the solutions of several transcendental equations, important for their
physical applications. All the aforementioned publications are available, with
open access, on Siewert's web page \cite{[Siewert-web]}; the symbols /S50,
/S52, etc., in the first lines of this paragraphs, indicate the number of the
respective paper in Siewert's publication list. The approach used in these
papers is based "on complex variable analysis and requires ultimately a
canonical solution of a certain Riemann problem; the solution of the suitably
posed Riemann problem follows immediately from the work of Muskhelishvili
\cite{[Mus]}", as stated in \cite{[S52]}. The effort invested in this vast
research is impressive, and the results are a pioneering and extremely
valuable contribution to the development of the theory of transcendental
equations. In the same time, the solutions obtained in this way are, in
general, very complicated and difficult to use in practical physical applications.

Recently, the interst for these solutions increased, as some of them can be
expressed in terms of generalized Lambert functions, and put in a much more
usable form, according to the results obtained by Mez\"{o}, Baricz
\cite{[Mezo]} and Mugnaini \cite{[Mugnaini]}. The applications of the theory
of generalized Lambert functions to various physical problems were presented
in \cite{[MezoKeady-arxiv]}, \cite{[MezoKeadyEJP]} and
\cite{[Barsan-arXiv2016]}.

From the point of view of applied physics, the efforts in getting approximate
analytical solutions to the same transcendental equations produced,
independently, useful results. The interference between the progress made in
mathematical physics and in applied mathematics (or in simple theories of
applied physics) was not discussed systematically, even if the subject seems
quite interesting. It is the main goal of the present paper to fill this gap.

So, author's intention was to interconnect results obtained in areas with a
small overlapping - mathematical physics, magnetism, quantum mechanics,
polymer physics, astronomy, solar energy conversion. The central contribution
of this paper is to point out to approximate solutions of Siewert's
transcendental equations and, whenever possible, to obtain approximate
expressions for generalized Lambert functions which describe these exact solutions.

The structure of this article is the following. In the second section, we
shall discuss a transcendental equation involving the Langevin function. Its
exact solution will be written in terms of a generalized Lambert function.
Using an analytical approximation of the inverse Langevin function, recently
proposed by Kr\"{o}ger, we find an approximate expression for this solution,
with a relative error smaller than $5\times10^{-3}.$ Such approximations are
useful not only in para- or super-paramagnetism, but also in polymer physics
and in solar energy conversion.

In Section 3, we shortly discuss two equations involving hyperbolic and
(linear) algebraic functions. The next one will be devoted to an equation
involving trigonometric and hyperbolic functions. Using an algebraic
approximation for the $\tan$ function, the solution of the transcendental
equation is written as a $W\left(  s;t;a\right)  $ generalized Lambert
function. An over-simplifying approximation of the hyperbolic function, of
interest for applied physics, is also mentioned. In Section 5, the asymptotic
solutions of two transcendental equations are expressed in terms of the Wright
$\omega$ function. In Section 6, several equations involving the Lambert and
generalized Lambert functions are mentioned, and in Section 7, transcendental
equations involving trigonometric and (linear) algebraic functions are
discussed. An approximate, quite precise solution of the Kepler equation for
elliptic orbits is discussed in detail. Section 8 is devoted to conclusions.

\section{The Langevin function and its inverse}

In \cite{[S80]}, Siewert and Burniston obtain an exact analytical solution of
the equation:%

\begin{equation}
x\coth x=\alpha x^{2}+1\label{1}%
\end{equation}

It can be written in terms of the Langevin function%

\begin{equation}
L\left(  x\right)  =\coth x-\frac{1}{x}\label{2}%
\end{equation}
as:%

\begin{equation}
L\left(  x\right)  =\alpha x\label{3}%
\end{equation}

\begin{figure}
\begin{center}
\includegraphics[width=\textwidth]{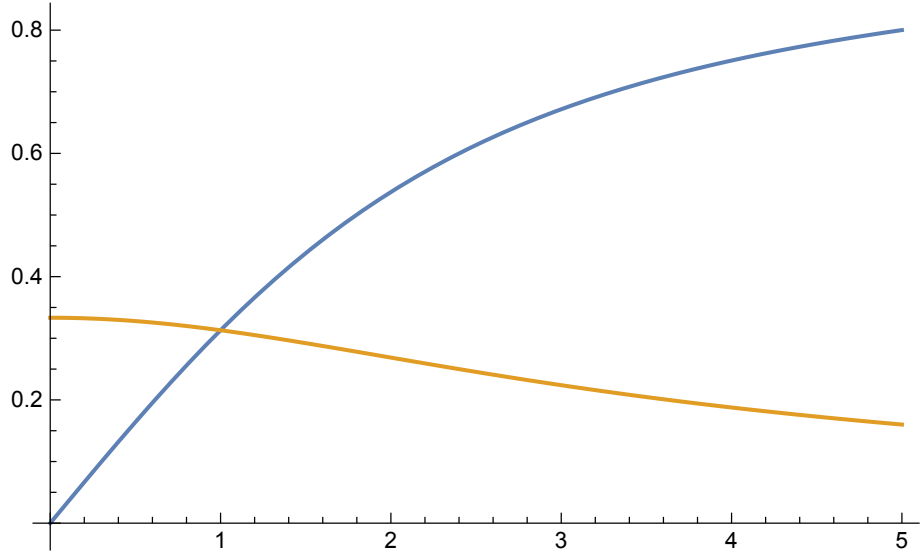}
\end{center}
\caption{The plots of L(x) (black) and L(x)/x (red).}
\end{figure}

It is easy to see that $L\left(  x\right)  $ is an odd, and $L\left(
x\right)  /x$ -\ an even function of $x.$

In order to express the solution $x\left(  \alpha\right)  $ of this equation
in\ terms of generalized Lambert functions, we shall put it in the form:%

\begin{equation}
e^{2x}\frac{\left(  \alpha x^{2}-x+1\right)  }{\left(  \alpha x^{2}%
+x+1\right)  }=1\label{4}%
\end{equation}
As:%

\begin{equation}
\alpha x^{2}-x+1=\alpha\left(  x-x_{1\alpha}\right)  \left(  x-x_{2\alpha
}\right) \label{5}%
\end{equation}

\begin{equation}
\alpha x^{2}+x+1=\alpha\left(  x+x_{1\alpha}\right)  \left(  x+x_{2\alpha
}\right) \label{6}%
\end{equation}
with:%

\begin{equation}
x_{1\alpha}=\frac{1+\sqrt{1-4\alpha}}{2\alpha},\ x_{2\alpha}=\frac
{1-\sqrt{1-4\alpha}}{2\alpha}\label{7}%
\end{equation}
we have:%

\begin{equation}
e^{2x}\frac{\left(  2x-2x_{1\alpha}\right)  \left(  2x-2x_{2\alpha}\right)
}{\left(  2x+2x_{1\alpha}\right)  \left(  2x+2x_{2\alpha}\right)  }=1\label{8}%
\end{equation}
so (\ref{4}) becomes:%

\begin{equation}
e^{2x}\frac{\left(  2x-t_{1}\right)  \left(  2x-t_{2}\right)  }{\left(
2x-s_{1}\right)  \left(  2x-s_{2}\right)  }=1\label{9}%
\end{equation}
with:%

\begin{equation}
t_{1}=2x_{1\alpha},\ t_{2}=2x_{2\alpha},\ s_{1}=-2x_{1\alpha}=-t_{1}%
,\ s_{2}=-2x_{2\alpha}=-t_{2}\label{10}%
\end{equation}
and its solution can be written as a generalized Lambert function:%

\begin{equation}
x\left(  \alpha\right)  =\frac{1}{2}W\left(  2x_{1\alpha},2x_{2\alpha
};-2x_{1\alpha},-2x_{2\alpha};1\right)  \ \label{11}%
\end{equation}

It seems that the value $\alpha=1/4$ plays no special role in the aspect of
the function $x\left(  \alpha\right)  ,$ even if the parameters $t_{1},\ t_{2}
$ are real for $\alpha<1/4$ and complex for $a>1/4.$

The Langevin function has been firstly introduced in the context of classical
theory of paramagnetism, where it gives the magnetization $M$\ as a function
of the external magnetic $H$\ field and temperature $T$:%

\begin{equation}
M=n\mu L\left(  \frac{\mu H}{k_{B}T}\right) \label{12}%
\end{equation}
(see for instance \cite{[Von]}, eq. (9.2)). This can be considered the
equation of state for a classical paramagnet. The same formula is valid for
superparamagnetic nanoparticles, at high enough values of temperature $T$
\cite{[wiki1]}, \cite{[KuncserSpringer]}.

The Langevin function is a particular case of the Brillouin function $B_{S},$
defined as:%

\begin{equation}
B_{S}\left(  x\right)  =\frac{2S+1}{2S}\coth\left(  \frac{2S+1}{2S}x\right)
-\frac{1}{2S}\coth\left(  \frac{1}{2S}x\right) \label{13}%
\end{equation}
Indeed,%

\begin{equation}
B_{\infty}\left(  x\right)  =L\left(  x\right) \label{14}%
\end{equation}
It is easy to see that, if $0<x<\infty,$ then:%

\begin{equation}
0<L\left(  x\right)  <1\label{15}%
\end{equation}
and:%

\begin{equation}
0<\frac{L\left(  x\right)  }{x}<\frac{1}{3}\label{16}%
\end{equation}

The Langevin function and its inverse are relevant not only for magnetism, but
also for other domains of physics with important practical applications, as
polymers (polymer deformation and flow) \cite{[Johal]}, \cite{[Kroger]},
\cite{[Jedynak]}, \cite{[Petrosyan]} or solar energy conversion (daily
clearness index) \cite{[Suehrcke]}, \cite{[Keady-Langevin]}. Researchers in
these fields proposed a large number of useful analytical approximations
for\ $L\left(  x\right)  $ and $L^{-1}\left(  x\right)  $. Less precise
algebraic approximations for $B_{S}\left(  x\right)  $ and $B_{S}^{-1}\left(
x\right)  ,$ but of real pedagogical interest,\ have been also obtained by
Arrott \cite{[Arrott]}. We shall exemplify the usefulness of such formulas in
the context of eq. (\ref{3}).

Taking the inverse Langevin function in both sides of (\ref{3}), we get:%

\begin{equation}
L^{-1}\left(  L\left(  x\right)  \right)  =L^{-1}\left(  \alpha x\right)
=x\label{17}%
\end{equation}
Let us use, for $L^{-1}\left(  x\right)  $, the very simple and precise
approximation proposed by Kr\"{o}ger, see eq. (10) of \cite{[Kroger]}:%

\begin{equation}
L^{-1}\left(  x\right)  =\frac{3x}{\left(  1-x^{2}\right)  \left(
1+0.5x^{2}\right)  }\label{18}%
\end{equation}
\qquad In this case, the transcendental equation%

\begin{equation}
L^{-1}\left(  \alpha x\right)  =x\label{19}%
\end{equation}
gives an approximate, but simple algebraic equation, whose physically
convenient root is:%

\begin{equation}
x\left(  \alpha\right)  =\frac{1}{\alpha}\sqrt{\frac{\sqrt{3\left(
3-8\alpha\right)  }-1}{2}}\label{20}%
\end{equation}

The identity%

\begin{equation}
f\left(  \alpha\right)  =\frac{x\left(  \alpha\right)  \coth x\left(
\alpha\right)  }{\alpha x\left(  \alpha\right)  ^{2}+1}=1\label{21}%
\end{equation}
where $x\left(  \alpha\right)  $ is replaced with the approximate solution
(\ref{20}), is fulfilled with a relative error less than $0.003$, as we can
see in the plot of Fig.2.

\begin{figure}
\begin{center}
\includegraphics[width=\textwidth]{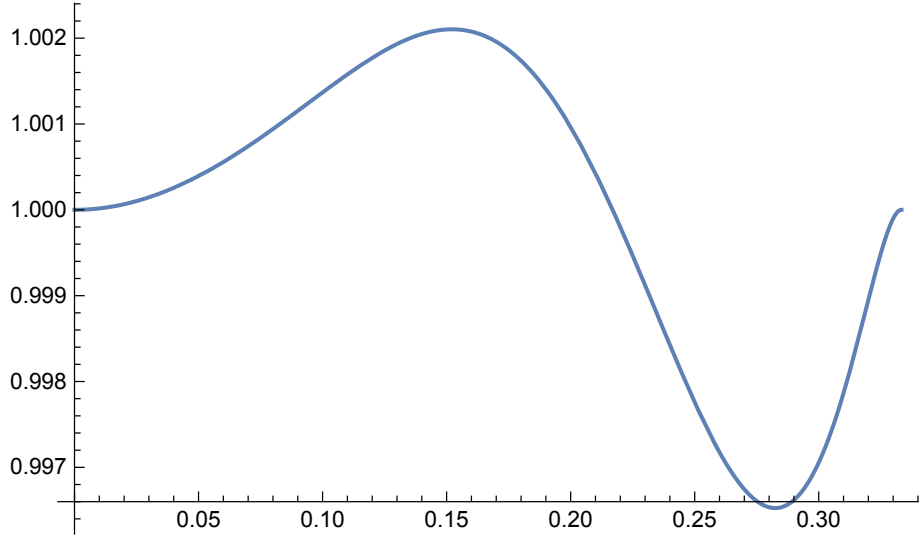}
\end{center}
\caption{The plots of f(a), eq. (21) and of the constant function y = 1.}
\end{figure}

So, we have the approximate relation:%

\begin{equation}
W\left(  \frac{1+\sqrt{1-4\alpha}}{\alpha},\frac{1-\sqrt{1-4\alpha}}{\alpha
};-\frac{1+\sqrt{1-4\alpha}}{\alpha},-\frac{1-\sqrt{1-4\alpha}}{\alpha
};1\right)  \simeq\label{22}%
\end{equation}

\[
\simeq\frac{1}{\alpha}\sqrt{2\left(  \sqrt{3\left(  3-8\alpha\right)
}-1\right)  }%
\]

The algebraic approximations for the Brillouin functions $B_{S},$ proposed by
Arrott \cite{[Arrott]}, are not very precise, but however very useful for
pedagogical purposes. The approach outlined in \cite{[Kroger]} might produce
much better results.

If we use Cohen's approximation, eq. (F3) of \cite{[Kroger]}:%

\begin{equation}
L^{-1}\left(  y\right)  =\frac{y\sqrt{3-y^{2}}}{1-y^{2}}\label{23}%
\end{equation}
we obtain, following the same steps%

\begin{equation}
y\left(  \alpha\right)  =\frac{1}{\sqrt{2}\alpha}\sqrt{\alpha\sqrt{\alpha
^{2}+8}-\alpha^{2}+2}\label{24}%
\end{equation}
which satisfies the identity%

\begin{equation}
g\left(  \alpha\right)  =\frac{y\left(  \alpha\right)  \coth y\left(
\alpha\right)  }{\alpha y\left(  \alpha\right)  ^{2}+1}=1\label{25}%
\end{equation}
- where $y\left(  \alpha\right)  $ is defined by (\ref{25}) - with a much
larger error compared to (21), as we can see in Fig. 3.

\begin{figure}
\begin{center}
\includegraphics[width=\textwidth]{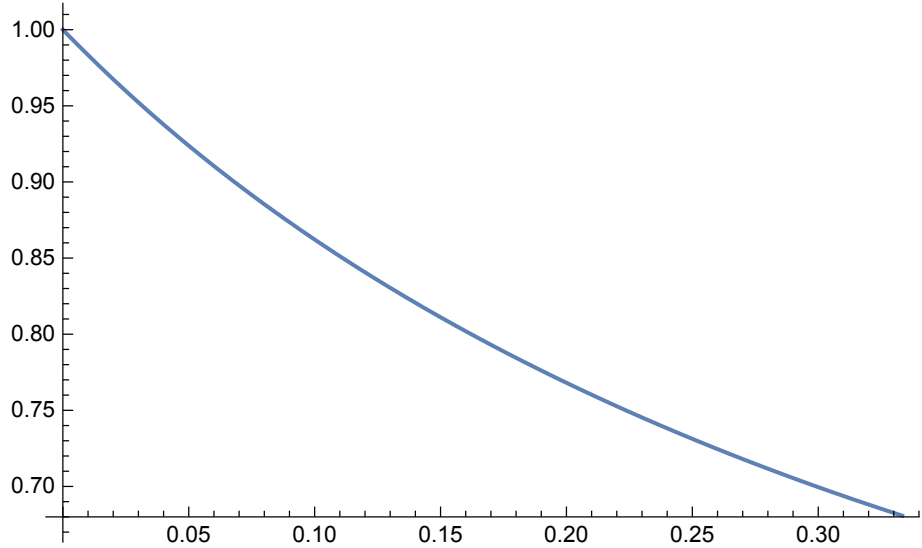}
\end{center}
\caption{The plots of g(a), eq. (25), and of the constant function y = 0.7.}
\end{figure}

\section{Equations involving hyperbolic and algebraic functions}

In \cite{[S63]}, the authors study the double zeros of the equation:%

\begin{equation}
x=\tanh\left(  ax+b\right) \label{26}%
\end{equation}

If $b=0,$ one obtains a numerical value for $a,$ namely $a=1,$\ equivalent to
the determination of the Curie temperature (see for instance \cite{[Stanley]},
eq. (6.15) or \cite{[Kittel]}, Ch. 15, eq. (8)).

Geometrically, a double zero of (\ref{26}) means that the line $y_{1}\left(
x\right)  =x$ is tangent to the curve $y_{2}\left(  x\right)  =\tanh\left(
ax+b\right)  ,$ consequently:%

\[
\frac{d}{dx}\tanh\left(  ax+b\right)  =\allowbreak-a\left(  \tanh^{2}\left(
b+ax\right)  -1\right)  =1
\]

But:%

\[
\tanh\left(  ax+b\right)  =x
\]
so the tangency condition becomes:%

\[
a\left(  1-x^{2}\right)  =1\rightarrow a=\frac{1}{1-x^{2}}%
\]
and the solution of the problem is reduced to solving eq. (2) of \cite{[S63]},
which can be written as:%

\begin{equation}
x=\tanh\left(  \frac{x}{1-x^{2}}+b\right)  =B_{1/2}\left(  \frac{x}{1-x^{2}%
}+b\right) \label{27}%
\end{equation}
We can see that, actually, the problem involves only one parameter, $b.$

The inverse of $\tanh$ can be expressed in terms of $\ln$, but it does not
produce a simpler equation. A precise algebraic approximation of these
functions could be of interest, as discussed at the end of the previous section.

In \cite{[S56]}, Siewert and Essig solve the Weiss equation of ferromagnetism:%

\begin{equation}
\zeta=\tanh\frac{1}{2}\left(  jz\zeta+h\right) \label{28}%
\end{equation}

Alternative ways of solving, exactly or approximately, this equation were
presented in \cite{[VB-VK]}, where the exact solution is written in terms of a
Lambert generalized function. The solution for the case $h=0$ was written as a
generalized Lambert function in  \cite{[MezoKeady-arxiv]},
\cite{[MezoKeadyEJP]}.

\section{Equations involving trigonometric and exponential functions}

In \cite{[S100]} and \cite{[S108]}, the authors obtain the solutions of an
equation basic to the theory of vibrating plates:%

\begin{equation}
a\tan x+\tanh x=0\label{29}%
\end{equation}
which appears also in quantum mechanics and electromagnetism.

We can made a certain progress in finding an approximate analytic solution of
this equation using the algebraic approximation fot $\tan x$ \cite{[dABG]}:%

\begin{equation}
\tan x\simeq\frac{0.45x}{1-\frac{2}{\pi}x},\ 0<x<\frac{\pi}{2}\label{30}%
\end{equation}
This formula can be easily extended for any real $x$ \cite{[BarsanPM2015]}.
Replacing $\tan x$ in (\ref{29}) according to (\ref{30}), we get:%

\begin{equation}
e^{2x}\frac{2x-\frac{\pi}{\left(  1-0.45a\frac{\pi}{2}\right)  }}{2x-\frac
{\pi}{\left(  1+0.45a\frac{\pi}{2}\right)  }}=1\label{31}%
\end{equation}

\begin{equation}
x\left(  a\right)  =\frac{1}{2}W\left(  \frac{\pi}{\left(  1-0.45a\frac{\pi
}{2}\right)  };\frac{\pi}{\left(  1-0.45a\frac{\pi}{2}\right)  };1\right)
\label{32}%
\end{equation}

Eq. (\ref{31}) is quite similar to the equation satisfied by the inverse
Langevin function:%

\begin{equation}
e^{2x}=\frac{A+1}{A-1}\frac{x+\frac{1}{A+1}}{x+\frac{1}{A-1}}\label{33}%
\end{equation}
so the recipes for obtaining $L^{-1}$ could be useful also for an approximate
evaluation of $W$ in (\ref{32}).

As $\tanh x$\ is a very slowly varying function, precise approximate solutions
for the $n-th~$root $\left(  n>1\right)  $ of (\ref{29}) can be obtained as
follows. Let us consider, for an illustrative example, that $a=-1.$

\begin{figure}
\begin{center}
\includegraphics[width=\textwidth]{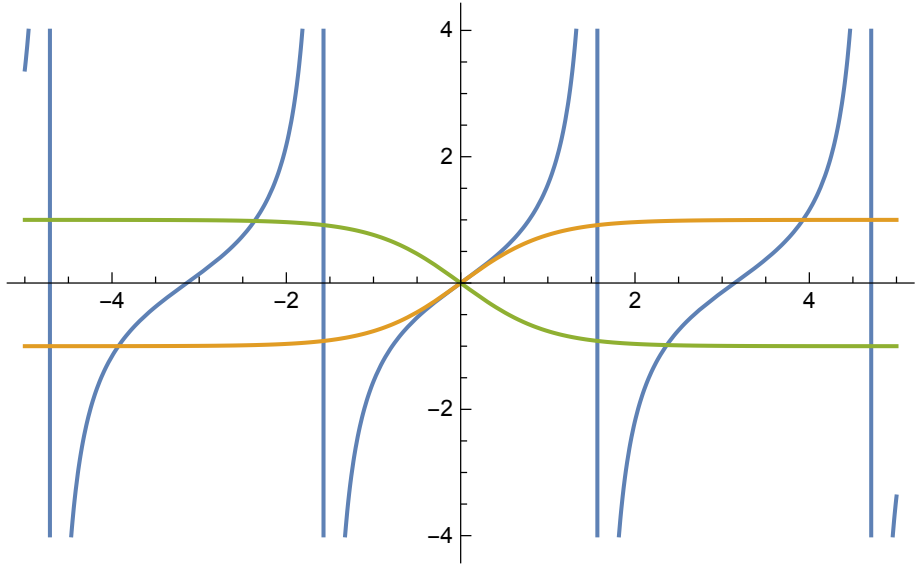}
\end{center}
\caption{The plots of tan(x) (black), tanh(x) (green) and - tanh(x) (red).}
\end{figure}

For the first root larger than $2\pi,\ $one can approximate $\tanh x$ as:%

\begin{equation}
\tanh x\simeq\tanh\left(  2\pi+\frac{\pi}{2}\right) \label{34}%
\end{equation}
and, putting $x=x_{0}+2\pi$ (i.e. reducing to the first quadrant):%

\begin{equation}
\tan\left(  x_{0}+2\pi\right)  =\tanh\left(  2\pi+\frac{\pi}{2}\right)
\label{35}%
\end{equation}
and%

\[
x_{0}=\arctan\left(  \tanh\left(  \frac{5\pi}{2}\right)  \right)  =0.785\,40
\]
so:%

\[
x=7.\,\allowbreak068\,6
\]

The "exact" value is:%

\begin{equation}
x=7.06858\label{36}%
\end{equation}
and the relative error:%

\begin{equation}
\varepsilon=\frac{7.06858-7.\,\allowbreak068\,6}{7.06858}=-2.\,\allowbreak
83\times10^{-6}\label{37}%
\end{equation}

So, for the practitioner working in applied physics, in a domain where the
experimental error is larger than $10^{-6},$ such a result is acceptable, for
pragmatic reasons.

The first root of the equation%

\begin{equation}
a\tan x+\tanh x=0\label{38}%
\end{equation}
can be obtained for small values of $a,~$after series expansions, as one of
the roots of the equation:%

\begin{equation}
\frac{17}{315}\left(  a-1\right)  x^{6}+\frac{2}{15}\left(  a+1\right)
x^{4}+\frac{1}{3}\left(  a-1\right)  x^{2}+a+1=0\label{39}%
\end{equation}
For instance, if $a=0.1,$ the error of the result obtained in this way is
about $-4\%$ \ ($x_{exact}=1.\,\allowbreak295\,2,\ x_{approx}=1.\,\allowbreak
343\,5$).

\section{The Wright omega function}

The Wright omega function appears in the asymptotic form of two equations
solved by Siewert and his co-workers.

In \cite{[S71]}, the authors obtain "an exact analytical solution for the
position-time relationship for an iverse-distance-squared force". Actually,
they study the repulsive classical 1D movement of an electric charge in the
field of another fixed charge. The repulsive force is given by the Coulomb law:%

\begin{equation}
m\frac{d^{2}r}{dt^{2}}=\frac{qQ}{4\pi\varepsilon_{0}r^{2}}\label{40}%
\end{equation}
The initial condition is:%

\begin{equation}
t=0\rightarrow\frac{dr}{dt}=0,\ r=r_{0}\label{41}%
\end{equation}
We shall define the position of the moving charge by the dimensionless
function $x\left(  t\right)  $, defined by:%

\begin{equation}
r\left(  t\right)  =r_{0}x\left(  t\right) \label{42}%
\end{equation}

After two integrations of the equation of movement, we get the relation
between position $x$\ and time $t:$%

\begin{equation}
\sqrt{x\left(  x-1\right)  }+\ln\left(  \sqrt{x}+\sqrt{x-1}\right)  =\frac
{t}{\tau}\label{43}%
\end{equation}
with $\tau$ given by:%

\begin{equation}
\tau=\sqrt{\frac{2\pi\varepsilon_{0}mr_{0}^{3}}{qQ}}\label{44}%
\end{equation}

We shall study this equation at small and at large values of $t.$ According to
(\ref{42}),%

\begin{equation}
x\left(  0\right)  =1\label{45}%
\end{equation}
so, for $t/\tau<<1,\ $we can put:%

\begin{equation}
x=1+X,\ \ X<<1\label{46}%
\end{equation}
and (\ref{43}) can be approximated by:%

\begin{equation}
\ln\left(  1+\frac{X}{2}+\sqrt{X}\right)  =\frac{t}{\tau}-\sqrt{X}\left(
1+\frac{X}{2}\right) \label{47}%
\end{equation}
and again, neglecting $X$ with respect to $\sqrt{X}$:%

\begin{equation}
\ln\left(  1+\sqrt{X}\right)  =\frac{t}{\tau}-\sqrt{X}\label{48}%
\end{equation}
or:%

\[
2\sqrt{X}=\frac{t}{\tau}%
\]
and finally:%

\begin{equation}
X=\frac{1}{4\tau^{2}}t^{2}\label{49}%
\end{equation}

At very small times, $t<<\tau,$ the movement is uniformly accelerated, as expected.

Asymptotically, $x>>1$ and (\ref{43}) gives:%

\begin{equation}
\ln\left(  2\sqrt{x}\right)  =\frac{t}{\tau}-x\label{50}%
\end{equation}
or:%

\begin{equation}
2xe^{2x}=\exp\left(  \frac{2t}{\tau}-\ln2\right)  \rightarrow2x=W\left(
\exp\left(  \frac{2t}{\tau}-\ln2\right)  \right) \label{51}%
\end{equation}
Consequently:%

\begin{equation}
x=\frac{1}{2}W\left(  \exp\left(  \frac{2t}{\tau}-\ln2\right)  \right)
\label{52}%
\end{equation}
where $W$ is the Lambert function. In terms of the Wright omega function
$\omega,$ we have the identity \cite{[CorlessW]}:%

\begin{equation}
W\left(  e^{x}\right)  =\omega\left(  x\right) \label{53}%
\end{equation}

So, the asymptotic formula (\ref{52}) can be written equivalently as:\qquad%

\begin{equation}
x\left(  t\right)  =\frac{1}{2}\omega\left(  \frac{2t}{\tau}-\ln2\right)
,\ \ t>>\tau\label{54}%
\end{equation}

The asymptotic expansion of the Lambert function is:%

\begin{equation}
W\left(  x\right)  =\ln x-\ln\left(  \ln x\right)  +\frac{\ln\left(  \ln
x\right)  }{\ln x}+...\tag{55}%
\end{equation}
Keeping only the first term, the asymptotic formula (\ref{52}) gives:%

\begin{equation}
x\left(  t\right)  =\frac{t}{\tau}-\frac{\ln2}{2}\simeq\frac{t}{\tau
}\label{56}%
\end{equation}

This is also an intuitive result, as, at very large distances, the repulsive
force produced by the fixed charge becomes negligable small, and the movement
becames almost uniform. So, the movement starts by being uniformly accelerated
and ends by being uniform.

In \cite{[S50]}, Siewert and Burniston solved the Kepler equation for
hyperbolic orbits:%

\begin{equation}
e\sinh F=F+N,\ \ N>0\label{57}%
\end{equation}
whose solution cannot be reduced to generalized Lambert function.
Asymptotically, $\sinh F\rightarrow\frac{1}{2}\exp F$ and (\ref{57}) becomes:%

\begin{equation}
\frac{e}{2}\exp F=F+N,\ \ N>0\label{58}%
\end{equation}
or, with $\exp F=f,\ F=\ln f:$%

\begin{equation}
\frac{e}{2}f=\ln f+N\ \label{59}%
\end{equation}
so, a Wright equation, whose standard form is \cite{[CorlessW]}:%

\begin{equation}
y+\ln y=z\label{60}%
\end{equation}

\section{Lambert function and generalized Lambert functions}

In \cite{[S62]}, Siewert and Burniston find the solution of the equation:%

\begin{equation}
ze^{z}=a\label{61}%
\end{equation}
i.e. obtain an expression for the $W$\ Lambert function, $a$ being a complex
parameter \cite{[CorlessL]}. In \cite{[S57]}, Siewert solves "the familiar
critical equation, described by age-diffusion theory, for a bare nuclear reactor":%

\begin{equation}
\frac{k\exp\left(  -B^{2}\tau\right)  }{1+B^{2}L^{2}}=1\label{62}%
\end{equation}
for $B^{2}$ (the buckling).

In \cite{[S68]}, the author solves a more compicated equation:%

\begin{equation}
e^{z}\frac{z}{z+b}=a\label{63}%
\end{equation}
with $a$ - a complex parameter. So, he obtains an expression for the function
$W\left(  0;-b;a\right)  $, which can be written, at its turn, in terms of the
Mez\"{o} - Baricz function $W_{r}.$ For $a$ - real, the author refers to a
paper of Wright, J. SIAM \textbf{9} (1961) 136.

\bigskip

\subsection{Transcendental equations involving trigonometric and algebraic
functions}

In \cite{[S59]}, the authors study "the critical condition for a spherical
reactor, described by elementary diffusion theory, surrounded by an infinite reflector":%

\begin{equation}
x\cot x=1-a-bx\label{64}%
\end{equation}
We can easily obtain an approximate analytical solution of (\ref{64}), using
the algebraic approximation of the tangent \cite{[dABG]},
\cite{[BarsanPM2015]}. We get, in this way, instead of (\ref{64}), an
approximate equation:%

\begin{equation}
x=\frac{0.45\pi\left(  x-n\pi\right)  }{2x-\left(  2n-1\right)  \pi}\left(
1-a-bx\right) \label{65}%
\end{equation}
which can be reduced to a second degree algebraic equation.

In \cite{[S53]}, Burniston and Siewert solve the equation:%

\begin{equation}
a\sin\zeta=\zeta\label{66}%
\end{equation}
which appears in quantum mechanics (defining the eigenenergies of a particle
in a square well potential), in electromagnetism, in elasticity, in optics
etc. Somewhat later, Siewert obtains a simpler solution \cite{[S89]}. A very
precise approximate analytic solution of (\ref{66}) was obtained through
algebraization \cite{[BarsanPM-SqWells]}; it is useful for the calculation of
energy levels in heterojunctions and quantum dots.

A more complicated variant of (\ref{66}) is the Kepler equation for elliptic
orbits \cite{[S50]} ($e$ is the excentricity and, in this section only, has
nothing to do with the basis of Nepperian logarithms):%

\begin{equation}
e\sin E=E-M,\ 0<e<1,\ 0<M<2\pi\label{67}%
\end{equation}

We can obtain a quite precise solution of (\ref{67}) approximating the first
half-bump of $\sin$ by a cubic polynomial:%

\begin{equation}
y\left(  x\right)  =ax^{3}+bx^{2}+x\label{68}%
\end{equation}
where the coefficients $a,~b~$\ can be determined by imposing the conditions:%

\begin{equation}
y\left(  \frac{\pi}{2}\right)  =1,\ y^{\prime}\left(  \frac{\pi}{2}\right)
=0\label{69}%
\end{equation}

We find:%

\begin{equation}
y\left(  x\right)  =\frac{4}{\pi^{3}}\left(  \pi-4\right)  x^{3}-\frac{4}%
{\pi^{2}}\left(  \pi-3\right)  x^{2}+x\label{70}%
\end{equation}
which fits quite well the function $\sin x,$ for $0<x<\pi/2,$\ as we can see
in Fig. 5.

\begin{figure}
\begin{center}
\includegraphics[width=\textwidth]{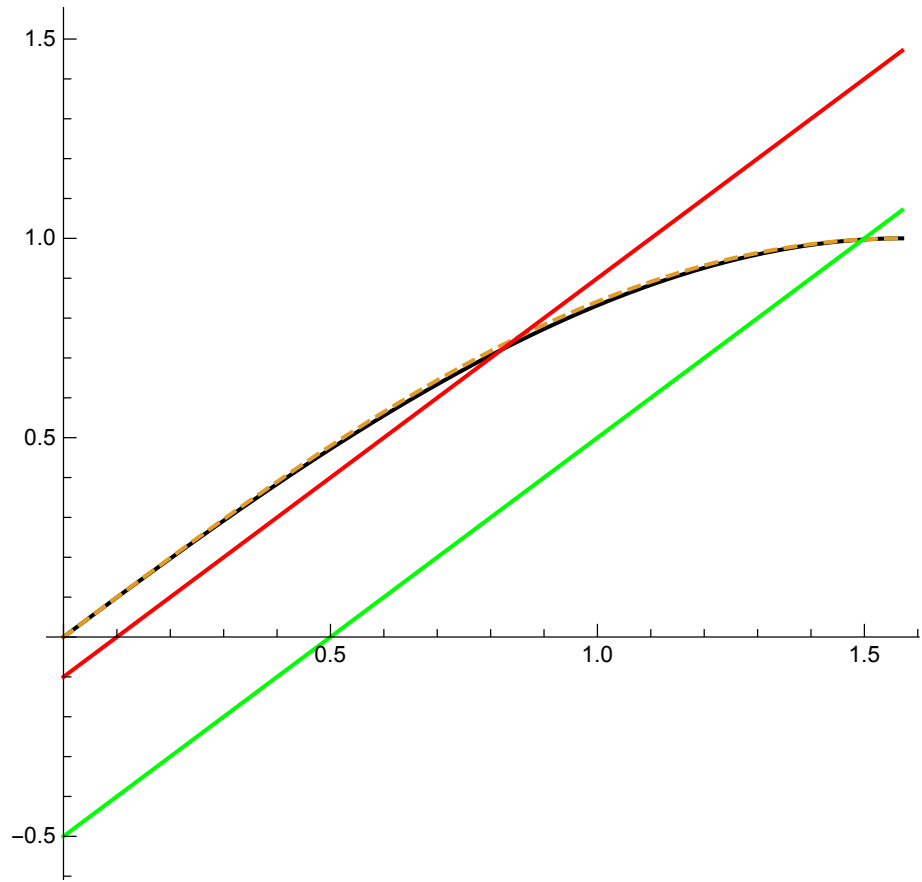}
\end{center}
\caption{The plots of y(x), eq. (70), black; sin(x), dashed; x - 0.1, red; x - 0.5, green.}
\end{figure}

If $e=0.9,\ M=0.1,$ replacing $\sin x$ with the polynom (\ref{70}) in the
Kepler equation (\ref{67}), we obtain the solution $x_{1p}=0.599\,55,$ while
the "exact" solution is $x_{1e}=0.599\,55;$ so, the error is $\varepsilon
_{1}\simeq5\times10^{-2}.$ If $e=0.9$ and $M=0.5,$ the polynomial
approximation gives $x_{2p}=1.\,\allowbreak382\,1,$ and the "exact" solution
is $x_{2e}=1.38441,$ so the error is$\ \varepsilon_{2}\simeq1.\,\allowbreak
7\times10^{-3}.$ The plots in Fig. 5 show, intuitively, why the second
approximation is more precise.

\section{Conclusions}

This paper is essentially focused on the transcendental equations studied by
Siewert and his coworkers, considered in conjunction with the results obtained
recently in the theory of generalized Lambert functions. Siewert's exact
results are compared, whenever possible, with the approximate analytical
solutions of the same equations, obtained with some simple techniques. Some
other results of Siewert and his coworkers, not connected to the generalized
Lambert functions, are discussed; in two cases, the asymptotic behavior of
Siewert's solutions are expressed in terms of Wright $\omega$ function.

As sometimes the approximate expressions of the generalized functions are very
precise (and their exact expressions are difficult to obtain), these
approximations could provide a useful guidance of their exact behaviour. Also,
the "algebraization" of the transcendental equations (i.e. the replacement of
the trigonometric functions with their various algebraic approximations) can
provide, sometimes, surprisingly precise analytic approximations. They can be
successfully used in applied physics or in the elementary presentation of
advanced problems.

\begin{acknowledgement}
\bigskip The author acknowledges the financial support of the IFIN-HH - ANCSI
project PN 16 42 01 01/2016 and to the IFIN-HH - JINR Dubna grant 04-4-1121-2015/17
\end{acknowledgement}

\bigskip

\bigskip

\bigskip

\end{document}